\documentclass[ams,preprint]{revtex4}
\usepackage{graphicx}
\usepackage{dcolumn}
\usepackage{amsmath}

\newcommand{\be}{\begin{equation}}
\newcommand{\ee}{\end{equation}}
\newcommand{\bea}{\begin{eqnarray}}
\newcommand{\eea}{\end{eqnarray}}

\begin{document}
\title{The Shift-Match Number and String Matching Probabilities for Binary Sequences}
\author{A. H. Bilge$^{1,2}$, A.
Erzan$^{2,3}$ and D. Balcan$^3$} \affiliation{$^1$\it Department
of Mathematics, Istanbul Technical University, Maslak 34469,
Istanbul, Turkey \\
$^2$ Feza G\"ursey Institute, P.O.B. 6, \c Cengelk\"oy, 34680
Istanbul, Turkey \\
$^3$\it Department of Physics, Istanbul
Technical University, Maslak 34469, Istanbul, Turkey}

\date{\today}

\begin{abstract}
We define the ``shift-match number'' for a binary string and we
compute the probability of occurrence of a given string as a
subsequence in longer strings in terms of its shift-match number.
We thus prove that the string matching probabilities depend not
only on the length of shorter strings, but also on the equivalence
class of the shorter string determined by its shift-match number.

\end{abstract}

\maketitle
\section{Introduction}
The sequence-matching problem can be defined as deciding whether a
given string over any alphabet occurs at least once as a
subsequence in another set of strings and it is a problem of
interest in informatics and genetics. From the informatics point
of view, the string-matching problem is essentially the
development of fast algorithms for determining exact or
approximate occurrences of a short string in longer strings. The
``words'' searched for are context and language dependent, hence
they should not be considered merely as a random selection from
the set of $n$-digit sequences of the given
alphabet~\cite{Fredriksson}. On the other hand  genomic
interactions~\cite{Sole} can be modelled as a valued graph, or
``network'' in which the nodes are strings of a given length. In a
previous set of papers, a model network has been proposed where
the  edges joining two nodes $(a,b)$ are placed according to
whether the string $a$ occurs {\it at least once} as subsequence
in the string $b$~\cite{Balcan,Mungan}. The number of nodes with a
given number of outgoing and incoming edges determine the
``out-degree'' and ``in-degree'' distributions, which require the
knowledge of the string matching probabilities for arbitrary
strings of given lengths. A closely related problem in biology is
the string alignment problem where it is important to determine
the probabilities of chance multiple occurrences of a random
string within a {\it specific} target string, allowing mismatches
and gaps~\cite{Karlin1,Karlin2}.

In the present paper we show that\ for an alphabet of length $2$,
i.e., binary strings, the matching probabilities of short strings
of length $n$ in {\it generic} longer strings of length $L$ fall
into equivalence classes with respect to a property of the short
strings which we identify as the ``shift-match number." We obtain
a recursive formula for the computation of the string matching
probabilities in terms of the ``shift-match number'' of the
shorter string, and the length $L$ of the longer strings. Counting
the total number occurrences (with multiplicities) of strings of
length $n$ in strings of length $L\ge n$, however, washes out the
fine structure induced by the shift-match number.

The paper is organized as follows. Section 2 is devoted to the
definition of the shift-match number, and proving the dependence
of the number of occurrences in longer strings to these numbers.
In Section 3 this is applied to the computation of the degree
distribution.

\section{The shift-match number}
\setcounter{equation}{0}

Let $a$ be a given binary string and let $P(a,L)$ be the
probability of occurrence of $a$ in binary strings of length $L$.
Since the number of distinct binary sequences of length $L$ is
$2^L$, this probability can be expressed as
\begin{equation}
P(a,L)=N(a,L) \; 2^{-L} \;,
\end{equation}
where $N(a,L)$ is the number of binary sequences of length $L$
that contain $a$ as a subsequence at least once. Clearly, if $a$
is a sequence of length  $n$, then $N(a,L)=0$ for $L<n$ and
$N(a,n)=1$. Furthermore the probability $P(a,L)$ approaches 1 as
$L$ increases.

We first computed $P(a,L)$ numerically for arbitrary $a$ of a
given length $n$ and for increasing values of $L$. We then
observed that although all sequences $a$ are equally likely to
occur, for fixed $L$, the probability $P(a,L)$ shows a variation
with respect to $a$ which reveals an equivalence class structure.
For example, for sequences of length $n=4$ we computed the
following probabilities $P(a,L)$, for $a$ as given in the first
line of the Table 1, and for $L=4,\dots, 10$.

Note that for $L=4$, the probabilities $P(a,4)$ are all equal to
$0.0625=2^{-4}$. For $L=5$, we see that the sequences fall into
two classes, with $a=1111$ being distinguished from the rest with
respect to its probability of inclusion, $P(a,5)$.  For $L=6$ and
$L=7$ we see that the equivalence classes branch further, into
three, then four. This structure stabilizes after $L=7$. We also
note that
\begin{eqnarray}
P(1000,L) = P(1100,L)&=& P(1110,L) > P(1001,L)=P(1011,L) \nonumber \\
&=&P(1101,L) >P(1010,L)> P(1111,L)
\end{eqnarray}
for each $L$. This observation was the starting point for our
definition of the ``shift-match number," which explains perfectly
the  equivalence class structure in the $P(a,L)$'s.

The computation of the number of occurrences $N(a,L)$ is motivated
by the counting algorithm (2.9) displayed in the proof of
Proposition 2.1. The final recursion formula (2.10) requires a
number of intermediate technical definitions such as the shift-
match indices $j_i$'s and $j_{i,m}$'s and the conditional number
of occurrences $N(a,L,m)$'s, to be defined in Section 2.2.

In dealing with binary numbers we use the following convention. If
$a$ is an $n$-digit binary number, the $i$'th digit of $a$ counted
from left is denoted by $a_i$, i.e.,
\begin{equation}
a=a_1a_2\dots a_n \; .
\end{equation}
We note that if $\bar{a}$ denotes the binary number obtained from
$a$ by replacing zeros with ones and vice versa, i.e., for
$a=00100110$, $\bar{a}=11011001$, then $P(a,L)=P(\bar{a},L)$.
Hence without loss of generality we may assume that $a_1=1$
wherever convenient. We now define the shift-match number $s(a)$
for a given binary number $a$ and the shift-match equivalence on
the set of binary numbers of a fixed length.

\subsection{The shift-match number and
shift-match equivalence}

Let $a$ be the $n$-digit binary number $a=a_1a_2a_3\dots a_n$. Its
shift-match number $s(a)$ is an $n$-digit binary number
$s(a)=s_1s_2s_3\dots s_n$ where the $s_i=s_i(a)$'s are defined by
\begin{eqnarray}
s_1(a) &=& 1 \; , \nonumber \\
s_2(a) &=& \delta(a_2,a_1) \; \delta(a_3,a_2) \; \dots
\; \delta(a_{n},a_{n-1}) \; , \nonumber \\
s_3(a) &=& \delta(a_3,a_1) \; \delta(a_4,a_2) \; \dots
\; \delta(a_{n},a_{n-2}) \; , \nonumber \\
     &\dots \nonumber \\
s_{n-1}(a) &=& \delta(a_{n-1},a_1) \; \delta(a_{n},a_{2}) \;, \nonumber \\
s_n(a) &=& \delta(a_n,a_1) \; .
\end{eqnarray}
The shift-match number induces an equivalence relation $\cong $ by
\begin{equation}
a \cong b \quad {\rm if \ and \ only \ if} \quad s(a)=s(b) \;.
\end{equation}
In the definition above, the choice   $s_1=1$ is a convention and
ensures that $s(a)$ and $a$ have same length. For $j\ge 2$,
$s_j=1$, if and only if the last $n-j+1$ digits of $a$ match with
its first $n-j+1$ digits.  We illustrate the computation of $s(a)$
by an example.  If $a= 1011011$, then by shifting $a$ to right
repeatedly, we obtain
\be
\begin{matrix}
a_1 & a_2 & a_3 & a_4 & a_5 & a_6 & a_7&\cr
    &     &     &     &     &     &    &\cr
1   & 0   & 1   & 1   & 0   & 1   & 1 &\quad \quad s_1=1 &\quad
({\rm by\ convention})\hfill&\cr
    & 1   & 0   & 1   & 1   & 0   & 1 &\quad \quad s_2=0 &\quad
 ({\rm no \ match\ with\ the\  first \ line})\hfill   &\cr
    &     & 1   & 0   & 1   & 1   & 0 &\quad \quad s_3=0 &\quad
 ({\rm no \ match\ with\ the\  first \ line})\hfill   &\cr
    &     &     & 1   & 0   & 1   & 1 &\quad \quad s_4=1 &\quad
    &\cr
    &     &     &     & 1   & 0   & 1 &\quad \quad s_5=0 &\quad
 ({\rm no \ match\ with\ the\  first \ line})\hfill   &\cr
    &     &     &     &     & 1   & 0 &\quad \quad s_6=0 &\quad
 ({\rm no \ match\ with\ the\  first \ line})\hfill   &\cr
    &     &     &     &     &     & 1 &\quad \quad s_7=1 &\quad
    & .
\end{matrix}
\nonumber \ee
Hence $s(a)=1001001$. We note that in many cases
$s(s(a))=s(a)$, i.e., the shift match number of $a$ is frequently
an element of the equivalence class of $a$, but this is not always
true, as for $a=11011$, $s(a)=10011$ and $s(s(a))=10001$, where
the latter does not belong to the equivalence class of $a$.

As we will show below that the probability of occurrence of $a$ in
longer strings will depend only on $s(a)$, we shall suppress   the
dependency of the shift-match number on $a$ and work with $s$. The
cardinality of the equivalence class whose shift-match number is
$s$  plays a crucial role in the computation of the network
connectivities and the corresponding probability will be denoted
by $P(s)$. We give in Table 2 the list of possible shift-match
numbers $s$ for binary sequences of length $n \le 6$ together with
the corresponding equivalence class ${\cal E} (s)$ and its
probability $P(s)$.

\subsection{The shift-match indices}
Recall that $a$ is a $n$ digit binary string starting with 1, $s =
s(a) = 1 \; s_2 \; \ldots \; s_n$ is the shift-match number of
$a$, and $s$ also is a $n$ digit binary number starting with 1. We
shall now give the definition of the shift-match indices $j_i$ and
$j_{i,m}$ for $1 \le i \le n$ and $1 \le m \le n-1$. We set
$j_1=1$, and define $j_i$ by
\begin{eqnarray}
j_1 &=& 1 \; , \nonumber \\
j_i &=& 2^{i-1} - \sum_{k=2}^i s_k \ j_{i+1-k} \quad {\rm for}
\quad i\ge 2 \; .
\end{eqnarray}
The number $j_i$ has the following meaning: write $a = a_1 \;
\ldots \; a_n$ in the first line, shift it to right once filling
with $\ast$ at the left and continue to get
\be
\begin{matrix}
a_1  & \dots  & \ldots & \dots  &\ldots & a_n \hfill &\cr
*    & a_1    & \dots  & \ldots &\ldots & a_{n-1} \hfill &\cr
*    & *      & a_1    & \dots  &\ldots & a_{n-2} \hfill &\cr
*    & \ldots & *      & a_1    & \dots & a_{n-i+1} \hfill &\cr
*    &\ldots  & \dots  &\ldots  & *     & a_1 \hfill &.
\end{matrix}
\nonumber
\ee
If we replace $\ast$'s with $0$'s or $1$'s, the $i$th line
represents $2^{i-1}$ distinct strings. If the shifted version of
$a$ has no match with the fist line, i.e., when $s = 1 0 \ldots
0$, then all strings are distinct. On the other hand when there
are matches, i.e., some of the $s_i$'s are non zero, there are
duplications, and the number of new strings added at each line
decreases. The indices $j_i$s are exactly the number of distinct
new strings added at the line $i$.

The ``conditional" shift-match indices $j_{i,m}$ for $1\le m \le
n-1$ are defined by
\begin{eqnarray}
j_{1,m} &=& s_{n-m+1} \; , \nonumber \\
j_{i,m} &=& s_{n-m+i} - \sum_{k=2}^i s_k \; j_{i+1-k,m} \quad {\rm
for} \quad 2\le i\le m  \; , \nonumber \\
j_{i,m} &=& 2^{i-m-1} - \sum_{k=2}^i s_k \; j_{i+1-k,m} \quad {\rm
for} \quad i\ge m+1 \; .
\end{eqnarray}
The numbers $j_{i,m}$ correspond to distinct sequences added at line
$i$, where first $m$ digits match with the last $m$ digits of $a$.
For example for $n=5$ we have
\be
\begin{matrix}
j_1=1 \hfill                 &j_{1,1}=s_n\hfill
&j_{1,2}=s_{n-1}\hfill\cr j_2=2-s_2 \ j_1 \hfill &j_{2,1}=1-s_2 \
j_{1,1}\hfill     &j_{2,2}=s_{n}-s_2 \ j_{1,2}\hfill\cr
j_3=2^2-s_2 \ j_2-s_3 \ j_1\;\;\;\; &j_{3,1}=2-s_2 \ j_{2,1}-s_3 \
j_{1,1}\;\;\;\; &j_{3,2}=1-s_2 \ j_{2,2}-s_3 \ j_{1,2}
\end{matrix} \nonumber
\ee
\noindent
These indices will be used in the computation of the string
matching probability for all short strings whose shift-match
number is $s$.

\subsection{The number of occurrences}
We define the conditional number of occurrences $N(a,L,m)$ for
$m=1,\dots n-1$, defined to be the number of times $a$ occurs as a
subsequence in sequences of length $L$ whose first $m$ digits
match with the last $m$ digits of $a$.

For example, if $a=110$, $L=4$, the sequences containing $a$ are
$0110$, $1100$, $1101$  and $1110$. Thus  $N(a,4)=4$, $N(a,4,1)$
is the number of such sequences starting with the last digit of
$a$ , i.e. $0$, hence $N(a,4,1)=1$. Similarly, $N(a,4,2)=0$, since
none of these starts with $10$.

For simplicity of the presentation we write \be j_{i,0} \equiv j_i
\; , \;\;\; N(a,L,0) \equiv N(a,L) \; . \nonumber \ee We shall now
give the expression of $N(a,L)$ in terms of $N(a,L-n)$ ,
$N(a,L-n,m)$ for $m = 1, \ldots, n-1$. This result shows, in
particular, that $P(a,L)$ depends only on the shift-match
equivalence class of $a$.

\vskip 0.3cm \noindent {\bf Proposition 2.1.} {\it Let $a$ be a
binary string of length $n$, $s$ be its shift-match number and the
shift-match indices $j_i$ and $j_{i,m}$ be defined by (2.6-7).
Then}
\begin{subequations}
\bea
N(a,L,m) &=& 0 \; , \; m=0, \dots, n-1 \quad {\rm for} \quad L<n \; , \\
N(a,n,m) &=& 1 \; , \; m=0,\dots, n-1,  \\
N(a,L,m) &=& \sum_{i=1}^{L-n+1} j_{i,m} \ 2^{L-n+1-i} \;, \;
m=0, \dots, n-1 \nonumber \\ &\quad& \quad \quad \quad \quad \quad \quad {\rm for} \quad n+1\le L< 2n-1 \; ,  \\
N(a,L,m) &=& \sum_{i=1}^n j_{i,m} \ 2^{L-n+1-i} + 2^{n-m} \ N(a,L-n,0) \nonumber \\
         &-& \sum_{i=1}^n j_{i,m} \ N(a,L-n,i-1) \; , \; m=0,\dots,n-1 \quad {\rm
for} \quad L\ge 2n-1 \;,  \\
N(a,L) &=& \sum_{i=1}^n j_{i} \ 2^{L-n+1-i} + 2^{n} \ N(a,L-n) \nonumber \\
       &-& \sum_{i=1}^n j_{i} \ N(a,L-n,i-1) \quad {\rm for} \quad L\ge 2n-1 \;.
\eea
\end{subequations}

\noindent {\bf Proof.}  We can enumerate the sequences of length
$L $ containing $a$ by counting the number of sequences of length
$L$ where the first occurrence of $a$ starts at the $k$th digit,
for $k=1,\dots , L-n+1$, as seen below.
\be
\begin{matrix}
a_1 &a_2 & \dots &a_{n-1} &a_n    & *  &  *  & \dots &\dots& * \cr
*   &a_1 & \dots &a_{n-2} &a_{n-1}&a_n &  *  & \dots&\dots& * \cr
*   &*   & \dots &a_{n-3} &a_{n-2}&a_{n-1}&a_n& \dots&\dots& * \cr
\dots &\dots&\dots&\dots  &\dots  &\dots
&\dots&\dots&\dots&\dots\cr
*   & *  & \dots & *      &a_1    &a_2 &a_3  &\dots    &\dots&*\cr
*   & *  & \dots & *      &*      &a_1 &a_2  &\dots    &\dots&*\cr
*   & *  & \dots & *      &*      & *  &a_1  &\dots    &\dots&*\cr
\dots &\dots&\dots&\dots  &\dots &\dots&\dots&\dots&\dots&\dots\cr
\dots &\dots&\dots&\dots &\dots  &\dots
&\dots&\dots&\dots&\dots\cr
*   & *  & \dots & *      &*      &*     &*    &\dots&\dots&a_n \cr
\end{matrix}
\ee
By assigning the values $0$ and $1$ to the $*$'s in the  first row
we obtain $2^{L-n}$ sequences containing $a$.  At the second row,
the same procedure gives again $2^{L-n}$ sequences but duplication
of the sequences already encountered in the first line should be
eliminated. It can be seen that sequences obtained from the first
and second rows cannot coincide unless the last and the first
$n-1$ digits of $a$ match, i.e., when $s_2=1$. Thus the number of
additional sequences is exactly $j_2\times 2^{L-n-1}$. By similar
considerations it can be seen that at the row $i$, the number of
additional sequences is $j_i\times 2^{L-n+1-i}$. For the
enumeration of the distinct sequences in the last $(L-2n+1)$'st
rows, we note that if there were no duplications with the ones in
the first $n$ rows, the contribution from this part would be
$2^m\times N(a,L-n)$. Duplications with the sequences already
enumerated at the $j$th row arise whenever the sequence of length
$L-n$ starts with the last $(n-i+1)$'st digits of $a$. But the
number of such sequences containing $a$ is what we call
$N(a,L-n,i-1)$, which is assumed to be known by the induction
step. Hence the proof is complete. \hfill $\bullet$

In Eqs.(2.8a-e), the number of occurrences are labelled by the
specific short string $a$, but it is clear from the proof that
they depend only on the shift-match number of $a$. In the next
section we shall use the notation \be N(s,L)=N(a,L,0) \;,
\nonumber \ee where $s=s(a)$ is the shift-match number of $a$.

\vskip 0.3cm \noindent {\bf Corollary 2.2.} {\it Let $s$ be a
shift-match number for binary string of length $n$. Then for each
member of the equivalence class ${\cal E}(s)$, the number of
occurrences in strings of length $L$ is given by}
\begin{subequations}
\bea
N(s,n) &=& 1 \; , \\
N(s,L) &=& \sum_{i=1}^{L-n+1} j_{i} \ 2^{L-n+1-i} \quad {\rm for}
\quad n+1 \le L< 2n-1 \;, \\
N(s,L) &=& \sum_{i=1}^n j_{i} \ 2^{L-n+1-i} + 2^{n} \ N(s,L-n) \nonumber \\
       &-& \sum_{i=1}^n j_{i} \ N(s,L-n,i-1) \quad {\rm for} \quad L\ge 2n-1 \;.
\eea
\end{subequations}

The probability of occurrence of any element of the corresponding
equivalence class is $P(s,L)=N(s,L)/2^L.$ The $P(s,L)$ for strings
of length $n=2,\dots 6$ in strings of length $L\le 12$ have been
computed using the formulas (2.10). These numbers have been
checked by direct enumeration of the numbers of matches of the
given strings in all strings of length $L$ using Octave. The
results obtained for the probability of occurrence of a given
string with shift-match number $s$ in a string of length $L$ are
displayed in Figure 1.

We display the numerical values of the $N(s,L)$ in Table 3,  where
$s$ denotes the shift match number of any representative of a
given equivalence class and $P(s)$ is the probability of this
equivalence class. The $L$ values appear at the top of the
respective columns. An inspection of Table 3 shows that for each
pair of lengths $n,L$, proceeding up the column of shift-match
numbers $s$, there exists a special value $s_c$ of $s$, such that
$N(s,L)=N(s_c,L)$ for all $s\le s_c$. Moreover, the total
probability $\sum_{s>s_c} P(s)$ of finding strings of length $n$
with shift match numbers larger than $s_c$ decreases rapidly with
$n$, as can be seen in Table 3. Thus the expected value $\langle
N(s,L) \rangle =\sum_s P(s) N(s,L) \to N(s_c,L)$ from below, as
$n$ becomes large, as shown in Table 4.

We will now use these findings to discuss the probability of
occurrence of a given string of length $n$ in a string of length
$L$, which is \be p(n,L) = \sum_s P(s) \ P(s,L) = \langle P(s,L)
\rangle \;. \ee In a previous paper by Mungan et
al.~\cite{Mungan}, an approximate expression has been given for
the probability $p(n,L)$, namely, \be p_M(n,L) \simeq
1-(1-2^n)^{L-n+1} \; . \ee The frequencies $\langle N(s,L)
\rangle$ reported in Table 4 have been obtained using Eqs.(2.10)
(first line for each $n$ value) and from $2^L p_M(n,L)$ (second
line for each $n$ value). A comparison of the numbers shows that
they are extremely close, which is gratifying since the
approximation (2.12) has been derived by a completely independent
route.

Furthermore, note that for $n$ and $L-n$ large, the expression in
Eq.(2.12) is approximately
\be p_M(n,L) \sim (L-n+1) /2^n \equiv
p_e(n,L) \; .
\ee
This corresponds to the limit where the
corrections to the probability, coming from correlations between
successive shifted sequences, can be neglected. In fact, we find
that $p_e(n,L) = N(s_c,L)/2^L$ exactly, or $N(s_c,L) =
2^{L-n}(L-n+1)$. It is once more instructive to note that in all
cases $ N(s_c,L) $ is the frequency associated with the smallest
shift-match numbers.

Had we asked for the total number of multiple occurrences of a
given string of length $n$ in a string of length $L\ge n$, we
would have found that the sum of these numbers over all sequences
of length $L$ depends only on $L-n$, and does not depend on the
shift-match number of the short sequence.  This total value
happens to be equal to the limiting values given in the last line
of Table 4, namely $N(s_c,L) = 2^{L-n}(L-n+1)$.

\section{Applications}
\setcounter{equation}{0}
Assume that we are given a collection ${\cal C}$ of strings of
lengths $L\le L_{\rm max}$ and let $N_L$ denote the number of
strings of length $L$. If $a$ is a binary string in this
collection, the degree of $a$, $d(a)$, is the number of sequences
in ${\cal C}$ that contain $a$ as a subsequence. The
considerations in Section 2 allow us to compute the degree
distribution in ${\cal C}$, over its elements $a$.

Define $d(s,L)$ as the expected number of occurrence of a string
with shift-match number $s$, in $N_L$ strings of length $L$,
\be
d(s,L) = P(s,L) \ N_L \; .
\ee
If the shift-match number of $a$ is
$s$, then the expected value of $d(a)$, depends only on $s$, and
we denote it by $\overline{d(s)}$. It can be seen that
\begin{eqnarray}
\overline{d(s)} &=& \sum_{L=n}^{L_{\rm max}} d(s,L) \; , \nonumber \\
\overline{d(s)} &=& \sum_{L=n}^{L_{\rm max}} P(s,L) \ N_L \; .
\end{eqnarray}
A brief inspection of Table 3 shows that the degree of a node
decreases with the string length $n$ and for fixed $n$, with the
shift-match number. Thus for fixed $L_{\rm max}$ we have the
ordering
\bea d(10)>d(11)>d(100)&>&d(101)>d(111) \nonumber
\\&>&d(1000)>d(1001)> d(1010)>d(1111)>\dots . \nonumber
\eea
As $N(s,L)$'s decrease with $s$ for fixed $L$, the
probabilities $P(s,L)$ also decrease with increasing $s$. Thus,
for any set of numbers $N_L$, the expected values
$\overline{d(s)}$ form a decreasing sequence. For lower values of
$s$ these numbers are strictly decreasing, but if the number
$N(s,L)$ and $N(s',L)$ coincide in the range of $n \le L \le
L_{\rm max}$ then $\overline{d(s)}$ and $\overline{d(s')}$ will be
the same, hence the equivalence classes $s$ and $s'$ will be
indistinguishable with respect to their string-matching
probabilities. We can view the computation of the string matching
probabilities on a collection of binary strings as a splitting
effect revealing a spectral structure.

In the numerical simulation of Ref.~\cite{Balcan}, faithfully
reproduced by the analytical approach of Refs.~\cite{Mungan}, the
average degree was computed for strings of length $n$, and the
degree distribution exhibited peaks centered at these average
values, with a certain variation about this value. We now see that
the degree distribution for strings of length $n$ in fact splits
into discrete spectral lines, identified with different
shift-match numbers. Thus, in the ideal case, where we can replace
$\overline{d(a)}$ by $\overline{d(s)}$, the spectrum is discrete.

The strength of the spectral lines correspond to the number of
strings $\nu(d(s))$ that have degree $d(s)$. If $s$ is the
shift-match number for the string of length $n$, then $\nu(d(s))$
is given by
\be \nu(\overline{d(s)}) =  \ P(s)\ N_n \; , \ee
where
$P(s)$ is the probability of the equivalence class as given in
Table 1 and $N_n$ is the total number of strings of length $n$. As
an example, for $s=100$ and $L_{\rm max}=6$ by using Table 3, we
easily get
\begin{eqnarray}
\overline{d(100)} &=& {1\over 8} N_3 + {4\over 16} N_4 + {12\over
32 } N_5 + {31\over 64} N_6 \; , \nonumber \\
\nu(\overline{d(100)}) &=& {2\over 4} N_3 \; . \nonumber
\end{eqnarray}
The results obtained for $\nu(\overline{d(s)})$ with respect to
$\overline{d(s)}$ have been shown in Figure 2, where we have
chosen $N_n= L_0p^2(1-p)^n$, for $p=0.05$ and $L_0 = 15000$, as in
Refs.[3,4].  Figure 2 may be compared to Figure 3 of
Ref.~\cite{Mungan}, where $\nu(d)$, averaged over 500 random
realisations of ${\cal C}$, has been plotted. Note that unless
$L_{\rm max}$ is large enough, the splitting between the values of
$\overline{d(s)}$ for successive $s$, does not show up, as can be
seen from the Table 3. Moreover, the $\overline{\nu(d(s))}$ values
are degenerate for different values of $s$, for small $n$.

The total number $q(n,L)$ of connections leading from strings of
length $n$ to strings of length $L\ge n$, averaged over 1000
random realisations, is also displayed in Table 5. These numbers
$q(n,L)$ should coincide with a weighted averages of $d(s)$'s over
shift-match numbers $s$'s. \be q(n,L) =  \sum_{|s|=n} \ P(s)\ N_n
\ N(s,L) \ 2^{-L} \ N_{L} \; , \ee where $|s|=n$ denotes the
sequences having shift-match number $s$ of length $n$. The numbers
computed from (3.4) agree with the numbers given in Table 5, given
that the numbers $N_n$ and $N_L$ are taken to be the number of
sequences of length $n$ and $L$ respectively, averaged over the
ensemble of realisations.

{\bf Acknowledgements}

AE would like to gratefully acknowledge a useful discussion with
Muhittin Mungan and partial support from the Turkish Academy of
Sciences.

\pagebreak

\begin{table}
\begin{ruledtabular}
\caption{The probabilities of occurrences of short strings $a$
 of length $n=4$ starting with $1$, in generic strings of length $L=4,\dots ,10$.}
\begin{tabular}{c|c|c|c|c|c|c|c|c}
{\boldmath $a$} & {\boldmath $1000$} & {\boldmath $1001$} &
{\boldmath $1010$} & {\boldmath $1011$}& {\boldmath $1100$} &
{\boldmath $1101$} & {\boldmath $1110$}&
{\boldmath $1111$}\\
\hline\hline
P(a,4) & 0.0625 & 0.0625 & 0.0625 & 0.0625 & 0.0625 & 0.0625 & 0.0625 & 0.0625 \\
P(a,5) & 0.1250 & 0.1250 & 0.1250 & 0.1250 & 0.1250 & 0.1250 & 0.1250 & 0.0938 \\
P(a,6) & 0.1875 & 0.1875 & 0.1719 & 0.1875 & 0.1875 & 0.1875 & 0.1875 & 0.1250 \\
P(a,7) & 0.2500 & 0.2422 & 0.2188 & 0.2422 & 0.2500 & 0.2422 & 0.2500 & 0.1563 \\
P(a,8) & 0.3086 & 0.2930 & 0.2656 & 0.2930 & 0.3086 & 0.2930 &
0.3086 & 0.1875 \\
P(a,9) & 0.3633 & 0.3398 & 0.3086 & 0.3398 & 0.3633 & 0.3398 &
0.3633 & 0.2168 \\
P(a,10) & 0.4141 & 0.3838 & 0.3486 & 0.3838 & 0.4141 & 0.3838 &
0.4141 & 0.2451
\end{tabular}
\end{ruledtabular}
\end{table}

\pagebreak

\begin{table}
\begin{ruledtabular}
\caption{List of the shift-match equivalence classes of binary
strings of length $ n \le 6$. The shift-match number is given in
the first column. Following the convention $a_1=1$, only half of
the elements in each equivalence class are displayed and they are
written in decimal form for compactness. The probability $P(s)$ is
obtained as the ratio of the cardinality of ${\cal E} (s)$ to
$2^n/2$.}
\begin{tabular}{c|c|c}
{\boldmath $s$} & {\boldmath $P(s)$} & {\boldmath ${\cal E} (s)$} \\
\hline\hline
10 & 1/2 & \{2\} \\
11 & 1/2 & \{3\} \\
\hline
100 & 2/4 & \{4,6\} \\
101 & 1/4 & \{5\} \\
111 & 1/4 & \{7\} \\
\hline
1000 & 3/8 & \{8,12,14\} \\
1001 & 3/8 & \{9,11,13\} \\
1010 & 1/8 & \{10\} \\
1111 & 1/8 & \{15\} \\
\hline
10000 & 6/16 & \{16,20,24,26,28,30\} \\
10001 & 5/16 & \{17,19,23,25,29\} \\
10010 & 2/16 & \{18,22\} \\
10101 & 1/16 & \{21\} \\
10011 & 1/16 & \{27\} \\
11111 & 1/16 & \{31\} \\
\hline
100000 & 10/32 & \{32,40,44,48,50,52,56,58,60,62\} \\
100001 & 11/32 & \{33,35,37,39,41,43,47,49,53,57,61\} \\
100010 & 3/32 & \{34,38,46\} \\
100011 & 3/32 & \{51,55,59\} \\
100100 & 2/32 & \{36,54\} \\
100101 & 1/32 & \{45\} \\
101010 & 1/32 & \{42\} \\
111111 & 1/32 & \{63\}
\end{tabular}
\end{ruledtabular}
\end{table}

\pagebreak

\begin{table}
\begin{ruledtabular}
\caption{In the table above the first column labelled by $s$ is
the shift-match number of the members of the equivalence class
while the second column labelled by  $P(s)$ is the probability of
occurrence of the corresponding equivalence class in strings of
length $n=2,\ldots, 6$. The numbers $N(s,L)$, of occurrences of
any string with shift-match number $s$ in sequences of length $L$,
are given in columns labelled by $L=2,\dots, 12$. Clearly nonzero
occurrences start after $L=n$. The numbers of occurrences decrease
as the shift match numbers increase. The probabilities of
occurrences, $P(s,L)$'s, are found by dividing the numbers in the
column labelled by $L$ with $2^L$.}
\begin{tabular}{c|c|ccccccccccc}
{\boldmath $s$} & {\boldmath $P(s)$} & 2  &3  & 4  & 5 & 6 & 7 & 8 & 9 & 10 & 11 & 12 \\
\hline\hline
10 & 1/2 & 1 & 4 & 11 & 26 & 57 & 120 & 247 & 502 & 1013 & 2036 & 4083 \\
11 & 1/2 & 1 & 3 & 8 & 19 & 43 & 94 & 201 & 423 & 880 & 1815 & 3719 \\
\hline
100 & 2/4 & & 1 & 4 & 12 & 31 & 74 & 168 & 369 & 792 & 1672 & 3487 \\
101 & 1/4 & & 1 & 4 & 11 & 27 & 63 & 142 & 312 & 673 & 1432 & 3015 \\
111 & 1/4 & & 1 & 3 & 8 & 20 & 47 & 107 & 238 & 520 & 1121 & 2391 \\
\hline
1000 & 3/8 & & & 1 & 4 & 12 & 32 & 79 & 186 & 424 & 944 & 2065 \\
1001 & 3/8 & & & 1 & 4 & 12 & 31 & 75 & 174 & 393 & 870 & 1897 \\
1010 & 1/8 & & & 1 & 4 & 11 & 28 & 68 & 158 & 357 & 792& 1731 \\
1111 & 1/8 & & & 1 & 3 & 8 & 20 & 48 & 111 & 251 & 558 & 1224 \\
\hline
10000 & 6/16 & & & & 1 & 4 & 12 & 32 & 80 & 191 & 442 & 1000 \\
10001 & 5/16 & & & & 1 & 4 & 12 & 32 & 79 & 187 & 430 & 968 \\
10010 & 2/16 & & & & 1 & 4 & 12 & 31 & 76 & 179 & 411 & 924 \\
10011 & 1/16 & & & & 1 & 4 & 12 & 31 & 75 & 175 & 399 & 894 \\
10101 & 1/16 & & & & 1 & 4 & 11 & 28 & 68 & 159 & 363 & 814 \\
11111 & 1/16 & & & & 1 & 3 & 8 & 20 & 48 & 112 & 255 & 571 \\
\hline
100000 & 10/32 & & & & & 1 & 4 & 12 & 32 & 80 & 192 & 447 \\
100001 & 11/32 & & & & & 1 & 4 & 12 & 32 & 80 & 191 & 443 \\
100010 & 3/32 & & & & & 1 & 4 & 12 & 32 & 79 & 188 & 435 \\
100011 & 3/32 & & & & & 1 & 4 & 12 & 32 & 79 & 187 & 431 \\
100100 & 2/32 & & & & & 1 & 4 & 12 & 31 & 76 & 180 & 419 \\
100101 & 1/32 & & & & & 1 & 4 & 12 & 31 & 76 & 179 & 412 \\
101010& 1/32 & & & & & 1 & 4 & 11 & 28 & 68 & 160 & 368 \\
111111 & 1/32 & & & & & 1 & 3 & 8 & 20 & 48 & 112 & 256
\end{tabular}
\end{ruledtabular}
\end{table}

\pagebreak

\begin{table}
\begin{ruledtabular}
\caption{In this table we present the values of $\langle
N(s,L)\rangle = \langle P(s,L) \rangle 2^L $, obtained in two
different ways. For each $n$, the first line corresponds to
Eq.(2.10), and the second line to the value obtained from
$p_M(n,L)2^L$ (Eq.(2.12)). See text.}
\begin{tabular}{c|cccccc}
{\boldmath $L-n$} & 0 & 1 & 2 & 3 & 4 & 5 \\
\hline\hline
n=2 & 1 & 3.5 & 9.5 & 22.5 & 50 & 107 \\
    & {1} & {3.5} & {9.25} & {21.875} & {48.8125} & {105.2188} \\
\hline
n=3 & 1 & 3.75 & 10.75 & 27.25 & 64.5 & 146.25 \\
    & {1} & {3.75} & {10.5625} & {26.4844} & {62.3476} & {141.1083}\\
\hline
n=4 & 1 & 3.875 & 11.375 & 29.625 & 72.25 & 168.625 \\
    & {1} & {3.875} & {11.2656} & {29.1230} & {70.6057} & {164.385} \\
\hline
n=5 & 1 & 3.9375 & 11.6875 & 30.8125 & 76.125 & 180.3125 \\
    & {1} & {5.9375} & {11.6289} & {30.5310} & {75.1538} & {177.6105} \\
\hline
n=6 &  1  & 3.96875 & 11.84375 & 31.40625 & 78.0625 & 186.15625 \\
    & {1} & {3.96875} & {11.8134} & {31.2577} & {77.5387} & {184.6544} \\
\hline Limit & 1  & 4 & 12 & 32 & 80 & 192
\end{tabular}
\end{ruledtabular}
\end{table}

\pagebreak

\begin{table}
\begin{ruledtabular}
\caption{In this table the average simulation results are given
for 1000 different realisations coming from Balcan-Erzan
model~\cite{Balcan}, assuming a length distribution $N_n= L_0 \
p^2 \ (1-p)^n$ for $p=0.05$ and $L_0=15000$. The numbers given
above show the average number of out going bonds from strings of
length $n$ (where $n$ is indicated in the first column) to strings
of length $L$ (first row).}
\begin{tabular}{c|c|c|c|c|c|c|c|c|c|c}
{\boldmath $(n,L)$} & 3  & 4  & 5 & 6 & 7 & 8 & 9 & 10 & 11 & 12 \\
\hline\hline
2 & 527.69 & 672.71 & 756.92 & 801.34 & 800.82 & 807.47 & 792.38 & 766.28 & 727.95 & 713.82 \\
3 & & 254.62 & 346.46 & 417.87 & 461.65 & 505.1 & 527.67 &
536.51 & 533.24 & 542.54 \\
4 & & & 117.47 & 163.89 & 200.12 & 233.59 & 259.75 & 279.06 &
288.77 & 305.76 \\
5 & & & & 53.35 & 74.82 & 94.6 & 111.53 & 124.75 & 133.84 &
146.89 \\
6 & & & & & 24.28 & 35.04 & 43.7 & 51.24 & 56.81 & 64.18
\end{tabular}
\end{ruledtabular}
\end{table}

\pagebreak

\begin{figure}
\leavevmode
\rotatebox{0}{\scalebox{1}{\includegraphics{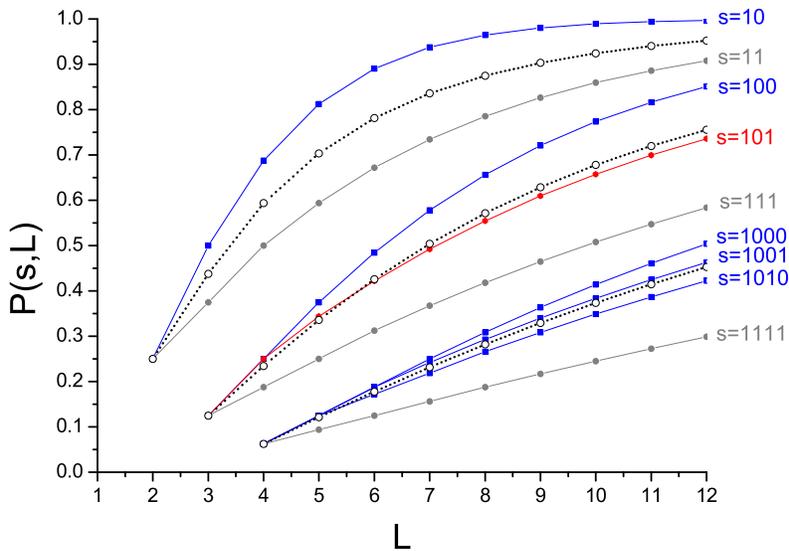}}}
\caption{The probability, $P(s,L)$, of a given string with
shift-match number $s$ to be reproduced in a randomly chosen
string of length $L$, plotted as a function of $L$. The dotted
line corresponds to the expectation value of $P(s,L)$ averaged
over the shift-match numbers $s$. It should be remarked that for
$n=3$, the branches for $s=100$ and $s=101$ are degenerate up to
$L=4$, where they split. For $n=4$, a first splitting occurs at
$L=5$ and a further ones at $L=6$ and $7$. See Table 3 for the
degeneracies in the number of occurrences $N(s,L)$, that give rise
to this progressive splitting. (Color online)}
\end{figure}

\pagebreak

\begin{figure}
\leavevmode
\rotatebox{0}{\scalebox{1}{\includegraphics{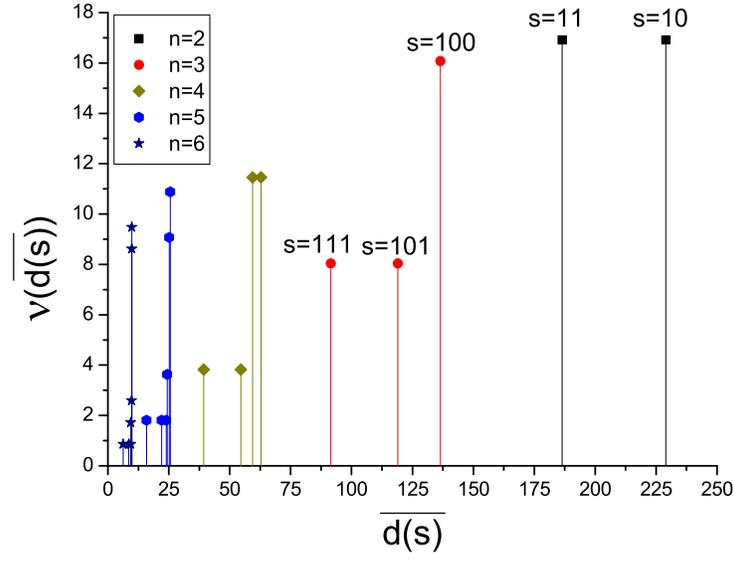}}}
\caption{The degree distribution $\nu(\overline{d(s)})$ of a given
string of length $n$ with shift-match number $s$ versus the degree
$\overline{d(s)}$ as computed from (3.3) and (3.2) respectively.
In the computation of $\nu(\overline{d(s)})$'s and
$\overline{d(s)}$'s the number of sequences of length $n$, $N_n$,
is taken to be $L_0 \ p^2 \ (1-p)^n$ for $p=0.05$ and $L_0 =
15000$, where the parameters have been chosen to facilitate
comparison with the numerical simulation of Ref.~\cite{Balcan}.}
\end{figure}

\end{document}